\documentstyle[12pt]{article}
\begin{document}
\thispagestyle{empty}
\def\cqkern#1#2#3{\copy255 \kern-#1\wd255 \vrule height #2\ht255 depth 
   #3\ht255 \kern#1\wd255}
\def\cqchoice#1#2#3#4{\mathchoice%
   {\setbox255\hbox{$\rm\displaystyle #1$}\cqkern{#2}{#3}{#4}}%
   {\setbox255\hbox{$\rm\textstyle #1$}\cqkern{#2}{#3}{#4}}%
   {\setbox255\hbox{$\rm\scriptstyle #1$}\cqkern{#2}{#3}{#4}}%
   {\setbox255\hbox{$\rm\scriptscriptstyle #1$}\cqkern{#2}{#3}{#4}}}
\def\CC{\mathord{\cqchoice{C}{0.65}{0.95}{-0.1}}}
\def\x{\stackrel{\otimes}{,}}
\def\y{\stackrel{\circ}{\scriptstyle\circ}}
\def\proof{\noindent Proof. \hfill \break}
\def\a{\begin{eqnarray}}
\def\b{\end{eqnarray}}
\def\p{{1\over{2\pi i}}}
\def\Q{{\scriptstyle Q}}
\def\P{{\scriptstyle P}}
\renewcommand{\thefootnote}{\fnsymbol{footnote}}

\newpage
\setcounter{page}{0}
\pagestyle{empty}
\centerline{\LARGE On ``Bosonic, Fermionic and Mixed" Super--}
\centerline{\LARGE symmetric $2$-Dimensional Integrable Models.}
\vspace{1truecm} \vskip0.5cm

\centerline{\large F. Toppan}
\vskip.5cm
\centerline{Dipartimento di Fisica}
\centerline{Universit\`{a} di Padova}
\centerline{Via Marzolo 8, I-35131 Padova}
\centerline{\em E-Mail: toppan@mvxpd5.pd.infn.it}
\vskip1.5cm
\centerline{\bf Abstract}
\vskip.5cm 
It is shown that supersymmetric integrable models in
two dimensions, both relativistic (i.e. super-Toda type 
theories) and non-relativistic (reductions of super-KP
hierarchies) can be associated to general Poisson-brackets 
structures given by superaffinizations of any bosonic Lie or any
super-Lie algebra.\par
This result allows enlarging the set of supersymmetric 
integrable models, which are no longer restricted to the  
subclass of superaffinizations of purely fermionic 
super-Lie algebras (that is
admitting fermionic simple roots only). 
~\par~\par
\pagestyle{plain}
\renewcommand{\thefootnote}{\arabic{footnote}}
\setcounter{footnote}{0}

{\bf Introduction}
\indent
Investigating the properties of two-dimensional integrable theories  
have become quite popular among high energy physicists during the last 
years. There are quite good reasons for that, among them we can mention 
the connections with string theory. It is clear by now that the 
non-relativistic integrable equations in ($1+1$) dimension of KdV or NLS
type encode the properties of the discretized version of two-dimensional
gravity (in single and multi-matrix models formulation).\par 
On the other 
hand two-dimensional relativistic theories of Toda type, whose simplest 
example is provided by the Liouville equation, are also relevant in many 
respect; for instance in the Polyakov formulation of string theory the
Liouville equation enters when dealing with non-critical strings. An 
even deeper connection results from the geometrical approach to string 
theory \cite{lr},\cite{lio}. This is related to and motivates some of 
the topics here 
discussed.
Since however they have been elucidated in the talk given by D. Sorokin, let 
me skip this point.\par
The problem of constructing supersymmetric generalizations of integrable models
is a very crucial one. The physical motivations are well-known, and even 
if no discretized version of super-Riemann surfaces leading to 
supermatrix models has been worked out so far, there is a hope that one 
can bypass this step assuming as fundamental objects the superintegrable
hierarchies themselves.\par
From a purely mathematical point of view the problem of classifying all
supersymmetric integrable models is quite challenging because 
of new features not present in the purely bosonic case. I will just 
mention here that in the bosonic case the situation is well-understood. 
Even if some problems are still opened (e.g. possible 
relations between hierarchies produced in different ways), the general
lines are clear: one starts with a given affine Lie algebra ${\cal G}$, 
then an integrable hierarchy can be produced either through hamiltonian
reduction \cite{bal} or through coset construction. This is true both 
for
non-relativistic hierarchies and for Toda models (in the latter case two 
copies of the affine Lie algebra should be taken, one for each 
chirality).\par
On the contrary the situation is different in the supersymmetric case; 
due to some simple argument which will be presented later, it was 
commonly believed \cite{hollo} that the only affine Lie algebra out of 
which one could obtain supersymmetric 
integrable hierarchies were the $N=1$ affinization
of the superLie algebras admitting a presentation in terms of fermionic 
simple roots only. For that reason only the integrable hierarchies
obtained from such affine superalgebras have been considered in the 
literature. \par
In this talk I will show that the above argument can be easily overcome 
and that interesting supersymmetric integrable models can be obtained
from $N=1$ affinizations of any bosonic Lie algebras, as well as any
super-Lie algebra (regardless if the simple roots are purely fermionic or 
necessarily some bosonic simple roots are present). With an abuse 
of language we can call the latter supersymmetric integrable models
either ``bosonic" or respectively ``mixed". ``Fermionic" supersymmetric 
integrable models are the previously known ones. Therefore
``bosonic", ``mixed" or ``fermionic" supersymmetries specify the sort of
(affine-Lie algebra) super-Poisson bracket structure we have to deal 
with. In any of these cases the resulting supermodels have ordinary
supersymmetric properties.\par~\par

{\bf The Matrix SuperKP Hierarchies} 
\indent
The starting point for a bosonic integrable hierarchy
in the AKS framework is a matrix-type Lax operator ${\cal L}$
\begin{eqnarray}
{\cal L} &=& {{\textstyle{\partial\over \partial x}}} +
J(x) +\Lambda
\label{lax}
\b
where $J(x)$ denotes a set of currents valued in the
semisimple finite Lie algebra ${\cal G}$. They give rise to an affine
algebra ${\hat{\cal G}}$ which provides (one of) the Poisson brackets 
structure of the underlying model. 
$\Lambda$ is a constant element in the loop algebra 
\a
{\tilde{\cal G}} &=& {\cal G} \otimes {\bf C} (\lambda, \lambda^{-1})
\b
where $\lambda$ is a spectral parameter. \par
If $\Lambda$ has a regularity property, that is if under its adjoint 
action ${\tilde{\cal G}}$ can be splitted into
\a
{\tilde{\cal G}} &=& {\tilde{\cal K}} \oplus {\tilde{\cal M}}
\b
where
\a
{\tilde {\cal K}} &=_{def}& Ker(ad_{\Lambda})\nonumber\\
{\tilde {\cal M}} &=_{def}& Im(ad_{\Lambda})
\b 
and ${\tilde {\cal K}}$ is abelian, while
\a
[{\tilde {\cal K}}, {\tilde {\cal M}}] &\subset& {\tilde {\cal M}}
\b
then, by a similarity transformation which is uniquely defined and 
iteratively computed order by order in negative powers of the spectral
parameter $\lambda$, we can diagonalize ${\cal L}\mapsto {\hat{\cal 
L}}$. ${\hat{\cal L}}$ is valued in the Cartan (abelian) subalgebra
of ${\cal G}$:
\a
{\hat{\cal L}} &=& \Lambda +\partial_x +J_\alpha h_\alpha +
\sum_{k=1}^\infty \lambda^{-k} R_{k,\alpha} h_\alpha
\label{laxd}
\b
The Cartan coefficients $R_{k,\alpha}$ are hamiltonian densities, 
whose integrals are in 
involution, for our integrable hierarchy, the Poisson brackets structure 
being given by ${\hat{\cal G}}$.\par
For a generic Lie algebra ${\cal G}$ there are many possible choices of
a regular element $\Lambda$ corresponding to different hierarchies, but 
for any Lie algebra at least two choices are always possible: {\it i)} 
$\Lambda$ is a sum over the simple positive roots of ${\tilde{\cal G}}$,
{\it ii)} $\Lambda$ is given by $\lambda H$ with $H$ any given Cartan 
element of ${\cal G}$. The first choice corresponds to generalized 
KdV-type hierarchies (KdV is recovered for $sl(2)$) while the second 
corresponds to generalized NLS-type hierarchies (standard NLS is 
obtained from 
$sl(2)$). \par
Inami and Kanno \cite{IK} proved that, under some restrictions,
the above construction can be
applied to the supersymmetric case. When dealing with $N=1$ 
supersymmetry one introduces a superspace parametrized by the bosonic and 
grassmann coordinate $x$, $\theta$ respectively and a 
fermionic derivative
\a
D \equiv D_X &=& {\partial\over \partial\theta} +\theta
{\partial\over \partial x}
\b
SuperKdV-type hierarchies can be produced from a matrix Lax operator
${\cal L}$ 
just as in the bosonic case, while ${\cal L}$ is now given by
\a
{\cal L} &=& D_X + \sum_i {\bf \Psi}(X) + \Lambda
\label{susylm}
\b   
Here ${\bf \Psi}(X)$ denotes $N=1$ supercurrents-valued on a (super-)Lie 
algebra whose Poisson brackets 
are the $N=1$ affinization of the given (super-)Lie algebra. Since we are 
dealing
with generalized KdV hierarchies, $\Lambda$ is given by the sum over the 
simple roots. It is already transparent from the above formula that, 
since $D$ and ${\bf \Psi}$ are fermionic, for consistency $\Lambda$ as 
well must be fermionic, which restricts the possible theories to those 
constructed from the superLie algebras which admit fermionic simple 
roots only. Indeed Inami and Kanno limited themselves to study this 
case.\par
The situation is clearly unsatisfactory, for instance one can ask what 
happens to supersymmetric NLS-type hierarchies: the regular element
$\Lambda$ should now be expressed by $\Lambda =\lambda H$ ($H$ in the 
Cartan) which is necessarily bosonic. The breaking of a definite 
statistics for the (eventual) ${\cal L}$ operator in this case 
apparently suggests that either supersymmetric extensions of 
NLS-type hierarchies do not exist or that these ones cannot be 
systematically produced via the AKS framework. Both these statements 
prove to be uncorrect. Indeed, by other means, it has been shown 
\cite{{brun},{top1}} that an integrable super-NLS hierarchy exists and moreover 
that admits as Poisson brackets structure the $N=1$ affinization of the 
(bosonic!) $sl(2)$ algebra.  Moreover it has a coset structure
w.r.t. the $N=1$ affine $U(1)$ subalgebra. This statement means that all 
the hamiltonian densities which provide the tower of hamiltonians in 
involution have vanishing Poisson brackets w.r.t. the above 
subalgebra.\par
At this point we have to understand if it is possible and how to fit 
such a result in the AKS framework. The ingredients have been given in 
\cite{top2}: it should be noticed the appearance of the spectral 
parameter $\lambda$ in $\Lambda = \lambda H$. 
Since the Lax operator ${\cal L}$ and 
its diagonalization are Laurent series in $\lambda$ it makes sense and 
is indeed possible to introduce the notion of ``alternated" or 
``twisted" bosonic or fermionic character of power series in $\lambda$,
\a
F(\lambda)  &=& \sum_{k=-\infty}^{+\infty} \lambda^{2k}(\xi_k + 
\lambda
\cdot \phi_k)
\b
is an ``alternated" fermion (boson) if $\xi_k$ are fermionic while 
$\phi_k$ are bosonic (and conversely). ``Alternated" fermions and bosons 
have the same ring properties as ordinary bosons and fermions. It is 
clear at this point that we can assume ${\cal L}$ being an 
``alternated" fermion and no contradiction with statistics will arise, for 
details see \cite{top2}. Notice that the theories produced out of this 
framework are ordinary supersymmetric theories in space and time since
$\lambda$ is only an auxiliary parameter.
~\par~\par
{\bf Hamiltonian reduction of any Super-WZNW model}\indent
For what concerns bosonic WZNW models, based on the Lie algebra ${\cal 
G}$, they are equivalent to two chiral copies $J$, ${\overline J}$ of 
the affine ${\hat {\cal G}}$ algebra
\a
J &=& \partial g\cdot g^{-1}\nonumber\\
{\overline J} &=& -g^{-1}\cdot {\overline \partial} g
\b
($g$ is valued in the group $G$ based on ${\cal G}$), satisfying the 
free equations od motion
\a
{\overline\partial} J= \partial{\overline J} =0&&
\b
The so-called abelian (hamiltonian) constrained bosonic WZNW
model is obtained by setting the positive (negative)
root component $J_>$ (${\overline J}_<$) to satisfy 
\a
J_> &=& \sum_i e_i\nonumber\\
{\overline J}_< &=& \sum_i f_i
\b
where the sums are over the positive (respectively negative)
simple roots of ${\cal G}$.\par
By inserting the Gau\ss~decomposition for $g$ the constrained model is
equivalent to a Toda field theory \cite{bal}. For $sl(2)$ we get the 
Liouville equation.\par
One could think to repeat the same steps in the $N=1$
supersymmetric case as 
well. As before we deal with a superspace, a fermionic derivative $D$, 
and fermionic supercurrents defined as
\a
\Psi &=& - i DG\cdot G^{-1}\nonumber\\
{\overline\Psi} &=& i G^{-1} DG
\b
with $G$ a supergroup element. \par
The free equations of the unconstrained model are
\a
{\overline D}\Psi = D{\overline\Psi} =0 &&
\b
Since $\Psi= \Psi_\alpha \tau^\alpha$ is fermionic, $\tau^\alpha$ are 
the generators of the (super-)Lie algebra ${\cal G}$, then
$\Psi_\alpha$ have opposite statistics w.r.t. the corresponding 
generator in ${\cal G}$. It follows that in order to repeat the 
same steps as before to constrain the theory, we need to have a 
superalgebra admitting fermionic simple roots only. So for instance the
standard superLiouville equation is recovered from the $osp(1|2)$ 
algebra which admits a single fermionic simple root. Moreover such 
constraints turn out to be superconformal and, after gauge-fixing, the 
Dirac's brackets provide a super-${\cal W}$ algebra (superVirasoro in 
case of $osp(1|2)$). For that reason it is commonly believed \cite{hollo}
that constraining superWZNW from bosonic algebras or ``mixed" 
superalgebras lead to non-supersymmetric models. Here again this 
statement proves wrong.\par
We can see this as follows \cite{st1}. Let us introduce 
a nilpotent Grassman differential
\a
d &=_{def}& (dz - i\theta d \theta ) \partial_z + d\theta D
\b
(it can be easily checked that $d^2=0$), we can introduce a Cartan form
\a
\Omega =_{def} dG \cdot G^{-1}
\b
which satisfy the Maurer-Cartan equation
\a
\relax d\Omega - {\textstyle{1\over 2}} [ \Omega , \Omega]_+ &=& 0
\b
where the anticommutator is understood in the Lie-algebraic context.  
\par
It follows that
\a
\Omega &=& (dz - i \theta d \theta ) J + i d\theta \Psi
\b
with
\a
J= J_{\alpha} \tau^{\alpha} &=_{def}& \partial G \cdot G^{-1}\nonumber\\
\Psi = \Psi_{\alpha}\tau^{\alpha} &=_{def}& -i DG\cdot G^{-1}
\b
As a consequence of the Maurer-Cartan equation
satisfied by $\Omega$ the $J_{\alpha}$ superfields are not 
independent,
but are constructed from the $\Psi_{\alpha}$ superfields:
\a
\relax J &=& D\Psi -{i\over 2} [\Psi , \Psi ]_+
\b
Let us specialize ourselves to the $sl(2)$ case (the most general case, 
along the same lines, is treated in \cite{st2}). We are now in the 
position to constraint the composite supercurrents $J_{\alpha}$ as 
before. Therefore we can set
\a
J_- &=& 1
\b
which allows us imposing a further gauge-fixing
\a
J_0|_{\theta=0} &=& 0
\b
Despite the fact that the above gauge-fixing is not manifestly
supersymmetric it turns out to be indeed superconformal, for details see
\cite{st1}.\par
The above constraint and gauge-fixing can be explicitly solved in terms
of the component fields entering the $\Psi_i$ superfields:
Let
\a
\Psi_i &=& \xi_i (z) + \theta j_i (z)
\b
(here $i=0,\pm)$.
In the $sl(2)$ case we are left with $3$ fundamental unconstrained
fields, two fermionic and one bosonic,
given by $\xi_-$, $ \xi_+$ and $j_+$, with spin
dimension respectively $-({1\over 2})$, ${3\over 2}$ and $ 2 $.\par
The remaining fields are expressed through these ones.\par
Performing the analogous constraint for the second chirality and 
reexpressing $\Psi$ through the superfields entering the 
Gau\ss~decomposition o f$G$ we are led with a superconformal
system of equations of motion
\begin{equation}\label{12}
\bar DD\Phi=e^{2\Phi}\bar\Psi^+\Psi^-, \qquad \bar
D\Psi^-=0=D\bar\Psi^+;
\end{equation}
\begin{equation}\label{13}
D\Psi^{-}+2D\Phi\Psi^-=1, \qquad
\bar D\bar\Psi^{+}+2\bar D\Phi\bar\Psi^+=1.
\end{equation} 
In component fields we are led with the following system:
\a
\Box\phi &=& e^{2\phi}\nonumber\\
{\overline\partial}\psi &=& 0\nonumber\\
{\partial{\overline\psi}}&=& 0
\b
with ${\psi}$, ${\overline\psi}$ free fermions and $\phi$ Liouville 
field. Such system as it can be checked is superconformal due 
to the nature of our 
constraints; the supersymmetry is realized non-linearly and 
spontaneously broken. Our system is based on a set of supersymmetric 
constraints. A peculiar feature is that the supersymmetric partner of a 
bosonic first class constraint is the second class. When analyzing the
Dirac's brackets of the surviving fields we can prove they are 
equivalent to a Virasoro (spin $2$ field) plus a free $b-c$ system of 
weight $(-{1\over 2},{3\over 2})$. The superconformal property of 
our model is reflected in the fact that there exists a Sugawara
realization of the superVirasoro algebra in terms of these fields.\par
The fact that bosonic and fermionic fields are decoupled is a peculiar
feature of the model based on $sl(2)$. It is not shared by more 
complicated models. In particular there is one which is rather interesting 
since it is 
based on the $osp(1|4)$ algebra. This is the simplest superalgebra
(the only at rank $2$, see \cite{sorba}) which admits a decomposition 
involving a simple fermionic and a simple bosonic root. 
This case is analyzed in 
\cite{st2}.
\par
Besides the nice mathematical properties of the above construction, the 
physical motivations are also quite important. More on that has been told 
by D. Sorokin.

\end{document}